\documentclass[%
reprint,
amsmath,amssymb,
aps,
pra,
floatfix
]{revtex4-2}

\usepackage{graphicx}% Include figure files
\usepackage{dcolumn}% Align table columns on decimal point
\usepackage{bm}% bold math
\usepackage{comment}
\usepackage{callouts}
\begin{document}
	
	\preprint{APS/123-QED}
	
	\title{Crystal fields, exchange, and dipolar interactions and noncollinear magnons of erbium oxide}% Force line breaks with \\
	
	\author{Kian Maleki}
	\author{Michael E. Flatt\'e}%
	
	\affiliation{%
		Department of Physics and Astronomy, University of Iowa, Iowa City, Iowa 52242, USA
	}%
	
	\date{\today}% It is always \today, today,
	
	\begin{abstract}
		We simulate the properties of magnons in erbium oxide,  a noncollinear antiferromagnet, from an effective single-ion Hamiltonian, including exchange and long-range dipolar interactions. 
        We parametrize the crystal field splitting of Er$_2$O$_3$ using Steven's operators and obtain the effective symmetry-dependent exchange constants between different erbium ions quenched by the crystal field at  different symmetry sites. We apply the Holstein-Primakoff transformation to the noncollinear spin system
        and  employ paraunitary diagonalization for the effective spin Hamiltonian. 
		The addition of the dipolar interaction to the exchange magnon dispersion changes the magnon bands drastically. The long-range nature of the dipolar interaction provides challenges to convergence, however we find that the averaged and normalized difference in the magnon dispersion is less than an averaged factor of $10^{-6}$  if the dipolar interaction is included out to the fortieth nearest neighbor.
       	\end{abstract}
	
	\maketitle

\section{Introduction}\label{sec: Intro}
Rare-earth ions embedded in solid state hosts play an important role in modern optical and quantum devices as they often manifest sharp optical lines and long optical and spin coherence times \cite{Spec_Prop_of_Rare_Earths}. They offer promise for the broadband quantum storage and transduction of photonic qubits required for secure communication between  future quantum computers \cite{Zhong2017}. Moreover, they are promising candidates for quantum information storage \cite{ZHANG2024112801}, quantum networks and other emerging technologies \cite{Ranci2018,Kolesov2020,Kimble2008,coccia2024converging,xiang2024molecular}.
Rare earth ions have also been used in single-molecule magnets which have potential applications in spintronics, information storage, and as potential qubits \cite{Bogani2008,202000090}. 
%In most electronic devices electrons move from one point to another and transfers information and energy. Technology takes advantage of the current and controls it with external electric fields and magnetic fields to create variety of devices. One candidate to replace the role of electron current is magnonic transfer of information and energy using magnons. 
Further potential spintronic applications include information transfer using magnons \cite{ZHENG2023129796,BHATT2025172275,PhysRevLett.132.193601}. Magnons are used for narrow band oscillators  \cite{8405247}, microwave  filters \cite{doi:10.1063/5.0075908}, and superconducting qubits \cite{doi:10.1126/science.aaa3693,doi:10.1126/sciadv.1603150,Lachance-Quirion_2019}. The long spin coherence times of rare-earth magnetic solids suggest the potential for long-distance magnonic transport such as seen in other materials \cite{YUAN20221,wang2024magnon,tengfei2025magnon,liang2025anisotropic}; although the magnetic transition temperatures of many rare-earth solids are far below room temperature the advent of cold electronic technologies for high-performance and quantum computing suggests potential utility \cite{SALIGRAM2024100082}. 
%Some of the reasons that makes magnons such an attractive option is that it has long lifetime as well as its available frequency and tunability .
%Magnons are very sensitive to magnetic fields and can be controlled and modified as one desires. 
After a single magnon was detected using a superconducting qubit \cite{doi:10.1126/science.aaz9236} several proposed quantum applications of magnons were explored, including qubit gates \cite{Candido_2021,PRXQuantum.2.040314,Ning:21,fukami2024magnon,Peng:25} and efficient quantum transducers \cite{Rochman2023,puel2024enhancementmicrowaveopticalspinbased}, which also do not require high operation temperature.

%Existence of edge state and the topological properties of magnons \cite{hu2024robust} is another interesting property of them.

%In 2021, It has been theoretically shown that two $NV$ (Nitrogen Vacancy) centers in diamond can be coupled via magnons \cite{PRXQuantum.2.040314}. 
%The magnon mediated entanglement is particularly interesting because magnons offer relatively long-range (a few $\mu m$) and spin coherent propagation and can also be coupled to each other \cite{tengfei2025magnon,liang2025anisotropic}. Some  hybrid quantum system use magnon to enhance the coupling between solid-state qubits \cite{Peng:25}.

Our focus here is on erbium oxide, as erbium has been used for several promising quantum devices. An erbium-doped crystal, placed in a microwave and optical resonator was used for quiet conversion of microwave photons to optical sideband photons with a potential for  100
\% quantum efficiency \cite{PhysRevLett.113.203601}. 
The long spin coherence times and optical accessibility at fiberoptic communications wavelengths of the electrons in Er$^{3+}$ have been used to propose and investigate  quantum memories, single photon sources, microwave to optical quantum transduction
\cite{ZhongGoldner+2019+2003+2015,doi:10.1126/sciadv.abj9786,Ranci2018,Huang:22,PhysRevB.101.184430,PhysRevB.105.245134} and transduction to a superconducting resonator \cite{PhysRevLett.110.157001}.	
%Another interesting feature of Er3+ emission is that it falls within the telecom C-band (1530 $nm$1565 $nm$) and that makes it  attractive for the application in solid-state quantum devices. \cite{10.1063/1.5142611,ZhongGoldner+2019+2003+2015}.
The intrinsic spin-photon interface and long coherence times of Er$^{3+}$ \cite{liu2006spectroscopic} make it suitable candidate for optical and microwave signal processing \cite{THIEL2011353,jiang2024quantum}.

Many appealing properties of the rare-earth elements originate  from the efficient shielding of electrons in  the $4f$ shell from surrounding perturbations by the $s$ and $p$ electrons of the ion. This electronic structure  can  lead to long spin coherence times for Er$^{3+}$ ions \cite{7010905}.  
The shielded $f$ electrons suggest  Er$_2$O$_3$ as a viable material host for both exchange and dipolar magnons. Here we develop a theory of the Er$_2$O$_3$ crystal field and its site symmetries,  as they play a complex role in  magnonic propagation. The site symmetries partially quench the orbital angular momentum despite the very strong spin-orbit interaction within these ions (Russell-Saunders coupling)  leading to a useful basis in total angular momentum %$J$ 
quantum numbers \cite{article_Rinehart}. We use the crystal field model to extract effective exchange constants between erbium ions with different site symmetry, spectral splittings and  effective Land\'e $g$~factors. We employ the molecular field approximation to find the dipolar interactions.  We calculate the magnon dispersion in the noncollinear antiferromagnetic ground state of the Er$_2$O$_3$ crystal using a generalized  Holstein-Primakoff (HP) transformation to describe the noncollinear systems. 
Finally, we study the effects of these parameters on the magnonic dispersion of Er$_2$O$_3$ and show that the dipolar interaction has a significant effect on the spectrum.

\section{Crystal field splittings and $g$ tensors of E\MakeLowercase{r}$_2$O$_3$} \label{sec: Crystal field}

\subsection{Crystal field splittings}

The general electron configuration of rare-earth atoms is ($n=11$ for erbium)
\begin{equation}
	\begin{split}
		{\rm La,\ Ce,\ Gd:} \ \ &(4f)^n (5s)^2(5p)^6(5d)^1(6s)^2,\\
		{\rm others:} \ \ &(4f)^{n+1} (5s)^2(5p)^6(6s)^2.
	\end{split}
\end{equation}
%where the first line corresponds to La, Ce, and Gd and the second line  to the rest of the rare-earth elements. 
In this electronic configuration the $4f$ electrons are shielded from outside perturbations including  the crystal field (CF) by complete $ (5s)^2 (5p)^6$ shells. The $(5d)^1(6s)^2$ or $(4f)^1(6s)^2 $ electrons  participate in bonding. Because the magnetic moment of the rare-earth materials comes from the remaining $4f$ electrons the magnetic moment of rare-earth ions does not change much from oxidation \cite{Gntherodt1976}. 

\begin{figure}
	% \begin{figure*} makes it one column 
		\includegraphics[width = 0.4 \textwidth]{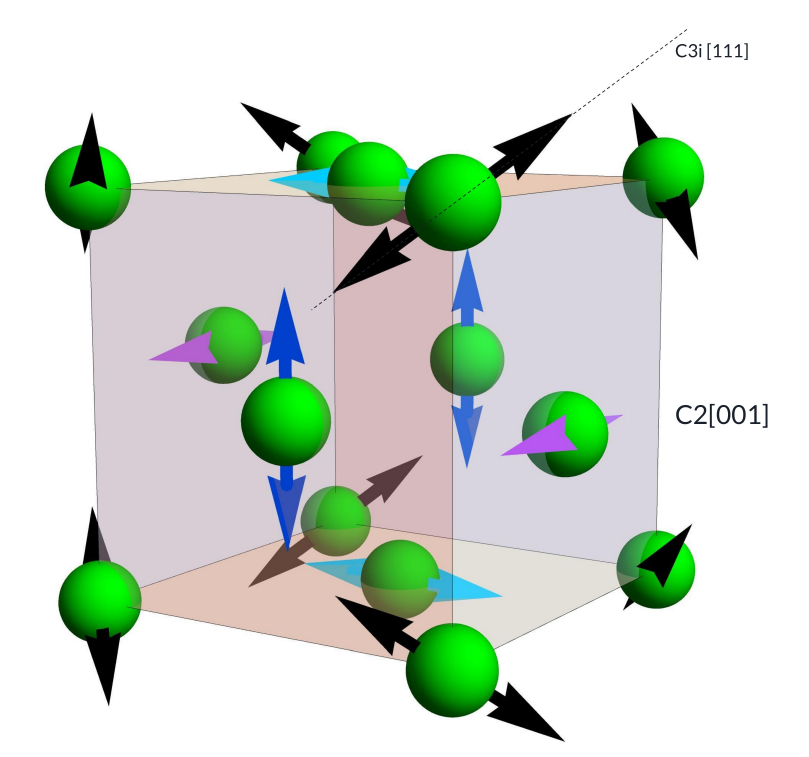}
		\caption{Schematic of all the $C_{3i}$ symmetry erbium ion sites, with symmetry axes, in a conventional (nonprimitive) cubic unit cell and six out of the the twenty-four $C_2$ symmetry erbium ion sites. The $C_{3i}$ sites are at the corners of a cube and the $C_2$ sites are in the faces of this cube. }
		\label{fig: er coor visualisation}
\end{figure}

In crystallized Er$_2$O$_3$ not all Er$^{3+}$ ions have the same site symmetry, see Fig. \ref{fig: er coor visualisation}. The  conventional (non-primitive) unit cell  has 32 erbium ions; 24  have a 2-fold rotational symmetry ($C_2$ sites) and the other 8 have a 3-fold rotational symmetry and inversion ($C_{3i}$ sites). The difference between these two different sites becomes apparent from the positions of the neighboring oxygen ions. At $C_{3i}$ sites, all six nearest oxygen atoms are 0.215372 $\text{\AA}$
away from the Er ions and the average position of these oxygen ions is co-located with the  Er ion position at the center.  At the $C_2$ sites, in contrast,  the distances of the oxygen ions from the erbium ion are  0.220095, 0.213963, and 0.211882$\text{\AA}$. The average position of the neighboring oxygen ions at each $C_2$ site is 0.014 $\text{\AA}$ displaced from the Er$^{3+}$ ion\cite{Ong2012b}.

The nearest erbium ion neighbors are 3.493 \r{A} apart, with next-nearest neighbors  3.55103 \r{A} apart. Since these are so close to each other we identify them both  as part of the first nearest neighbor shell of erbium ions. Similar situations happen for the second (located at 3.9835 or 3.9987 \r{A}) and the third (5.2752 or 5.3209 \r{A}) nearest neighbors. 

Group theory can be used to show that in case of cubic symmetry, the energies of a $15/2$ total angular momentum state split into doublets ($\Gamma_6$ and $\Gamma_7$) as well as three quartets ($\Gamma_8$),
\begin{equation} 
	\begin{aligned}
		D^{15/2} = \Gamma_6 \oplus \Gamma_7 \oplus 3 \Gamma_8
	\end{aligned} 
\end{equation} 
where $\Gamma_6$ and $\Gamma_7$ are two-dimensional representations and $\Gamma_8$ is a four-dimensional representation. For lower site symmetries such as $C_2$ and $C_{3i}$ the three $\Gamma_8$'s each split into two dimensional representations, making a total of six. The ground state can be any of these representations. As shown in the next section, the ground state depends on the  parameters of the model. For more details about the relation between the fit parameters of this model and of the ground state see \cite{Ammerlaan2001-kp}.

\begin{figure*}[!ht]
	% \begin{figure*} makes it one column 
		\includegraphics[width = 0.95\textwidth]{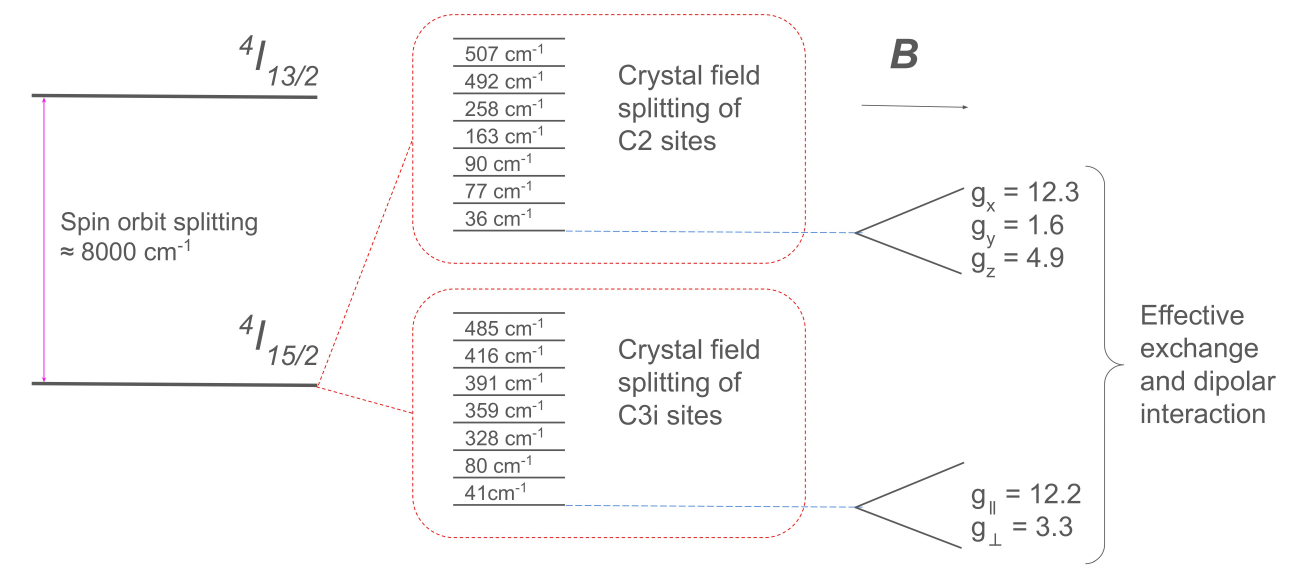}
		\caption{Energy splitting of the Er$^{3+}$ ions in Er$_2$O$_3$. The ground state multiplet $^4I_{15/2}$ is split into eight Kramers doublets. The crystal field splitting differs for $C_2$ and $C_{3i}$ sites. The ground states of the crystal field are used to find the effective $g$~factors, exchange interactions and dipolar constants\cite{Dammak_200301822,Ammerlaan2001-kp}.}
		\label{fig: crystal field splitting diagarm}
\end{figure*}

The strong spin-orbit interaction within each  Er ion creates energy splittings of $\sim$800~meV$\sim$200~THz$\sim$8000~cm$^{-1}$, whereas the crystal field is about $\sim$50~meV$\sim$12 THz$\sim$650~cm$^{-1}$. The magnons at very low temperature are dominated by the lowest Kramers doublet, see Fig. \ref{fig: crystal field splitting diagarm}.

In order to model the crystal field Hamiltonian we use  extended Stevens operators ($O^q_k$) \cite{laachir_crystal-field_2011,10.1063/1.349382}. Each Stevens operator is a function of $S^x, \; S^y ,$  and/or $ S^z$.
Note that here $S$ is the total angular momentum, used to avoid confusion because $J$ will refer to the  exchange constants. We construct a linear combination of the Stevens operators such that it has the desired symmetry. We focus on the crystal field Hamiltonian for each site symmetry and find the fit parameters that yield the correct crystal field splitting energies and $g$~factors. 
Similar work on Er$_2$O$_3$ was done to fit the CF energy splitting \cite{Dammak_200301822} or the $g$~factors  \cite{Ammerlaan2001-kp}. Here we find the parameters that give the best match to both the CF energy splittings and $g$~factors.

For a cubic symmetry the combination of the Stevens operators is
	\begin{equation} 
		\begin{aligned}
			&H_4 = 5 O^4_4 + O^0_4\\
			&H_6 = -21 O^4_6 + O_6^0 \\
			&H_{cube} =  B_4 H_4 + B_6 H_6
		\end{aligned} 
	\end{equation}
where $B_4$ and $B_6$ are fit parameters. 
Adding the following terms to the cubic Hamiltonian yields a crystal field Hamiltonian with $C_2$ symmetry,
\begin{equation} 
	\begin{aligned}
		&H_{te} = B_{te} O_2^0 \\
		&H_{or} = B_{or} O_2^2 \\
		&H_{C_{2i}} = B_{2i} O_{2}^{-2} 
	\end{aligned} 
\end{equation}
where $B_{te}$ (tetragonal symmetry), $B_{or}$ (orthorhombic symmetry), and $B_{2i}$ are again fit parameters. For Stevens Operators the quantization axis is usually chosen to be $\hat{z}$. %This is can be checked by looking at $O_2^0 = 3 S_z^2 -s \mathbb{I}$, since only the quantization axis appears in this term. 
In order to get $H^{CF}_{C_{2x}}$ and $H^{CF}_{C_{2y}}$ we perform  $SO(3)$ operations on the Stevens operators , $S_x \rightarrow S_y \rightarrow S_z \rightarrow S_x$ and $ S_x \rightarrow S_z \rightarrow S_y \rightarrow S_x $ , respectively. Performing the symmetry operators does not change the crystal field splitting but it rotates the $g$~tensor around.

So for $C_2$ we obtain
\begin{equation} 
	\begin{aligned}
		H^{CF}_{C_{2z}} = H_{4} + H_{te} + H_{or} + H_{C_{2i}}
	\end{aligned} 
\end{equation}
\noindent which requires five fit parameters. We used Monte Carlo $\chi^2$  minimization to find the best values for these parameters, which are listed in  Table~\ref{tab: Paremeters of Steves operators}. The energy splittings and the $g$~factors resulting from these crystal field parameters are in Table~\ref{tab: $C_2$ splitting value}.

For $C_{3i}$ sites	with a symmetry axis along $[111]$, first we define a trigonal term and then combine it with the cubic term as follows
\begin{equation} 
	\begin{aligned}
		&H_{tri} = \frac{1}{2} O_2^{-2} + O_2^1 + O_2^{-1}\\
		&H_{C_{3i[111]}} = H_{cube} + B_{[111]} H_{tri}
	\end{aligned} 
\end{equation}
where $ B_{[111]}$ is the fit parameter. This requires three fit parameters. For the other $C_{3i}$ sites we have
  $[\bar{1}11]$, $[1\bar{1}1]$, and $[11\bar{1}]$  corresponding to $S_x \rightarrow -S_x$, $S_y \rightarrow -S_y$, and $S_z \rightarrow -S_z$, respectively. The resulting parameters are shown in Table~\ref{tab: Paremeters of Steves operators}.  The energy splittings and the $g$~factors are  in Table~\ref{tab: $C_{3i}$ splitting values}.

%%%  unit conversion: 1 cm-1 = 0.124 meV
\begin{table}
	\caption{Best fit parameters to produce the  energy splittings and $g$~factors}%
	\begin{ruledtabular}
		\begin{tabular}{cll}
			\label{tab: Paremeters of Steves operators}
			\textrm{Symmetry type}&
			\textrm{Parameter}&
			%\textrm{Value ($meV$)}&
			\textrm{Value (cm$^{-1}$)} \\
			\hline
            \hline
			& $B_4$& $2.48 \times 10^{-3}$\\
			& $B_6$& $-2.843 \times 10^{-5} $\\
			$C_2$& $B_{te}$ &   $0.533$\\
			& $B_{or}$&  $-1.098$\\
			& $B_{2z}$&  $-4.038$\\
			\hline
			 & $B_4$&   $1.24 \times10^{-2}$\\
			$C_{3i}$ & $B_6$&  $7.525 \times 10^{-6}$\\
			 & $B_{[111]}$ &     $-3.89$
		\end{tabular}
	\end{ruledtabular}
\end{table}

\subsection{$g$ tensors}

In the absence of external magnetic fields the lowest energy eigenstates 
of the crystal field Hamiltonian, $\mathcal{H}_{CF}$, are degenerate with eigenvectors $|v_1\rangle$ and $|v_2\rangle$. The Zeeman Hamiltonian is%,
\begin{equation}
	\mathcal{H}_{z} = g_L \mu_B \mathbf{B}\cdot\mathbf{S},
\end{equation}
where $\mathbf{S}$ (total angular momentum of $15/2$) is a $16 \times 16$ matrix, and $g_L=6/5 $ is the Land\'e $g$~factor. Projecting the Zeeman Hamiltonian onto the lowest eigenstates of $\mathcal{H}_{CF}$, we obtain
\begin{equation}\label{eq: 2x2 Zeeman Hamiltonian of 16x16 spin operator}
	\mathcal{H}_{z} = g_L \mu_B 
		\begin{pmatrix}
		\langle v1 | \mathbf{B}\cdot\mathbf{S} |v1 \rangle & \langle v2| \mathbf{B}\cdot\mathbf{S} |v1\rangle \\
		\langle v1| \mathbf{B}\cdot\mathbf{S} |v2\rangle	 &	\langle v2| \mathbf{B}\cdot\mathbf{S} |v2\rangle 
	\end{pmatrix}
\end{equation}
We would like to reduce the dimension of the relevant Hilbert space associated with this Hamiltonian. This can be achieved by comparing it with the Zeeman Hamiltonian of a spin 1/2 particle.
\begin{equation} \label{eq: 2x2 Zeeman Hamiltonian of effective spin}
	\mathcal{H}_{z} = g_{eff} \mu_B 
	\begin{pmatrix}
		\langle \uparrow | \mathbf{B}\cdot\mathbf{s} |\uparrow \rangle & \langle \downarrow| \mathbf{B}\cdot\mathbf{s} |\uparrow\rangle \\
		\langle \uparrow| \mathbf{B}\cdot\mathbf{s} |\downarrow\rangle	 &	\langle \downarrow| \mathbf{B}\cdot\mathbf{s} |\downarrow\rangle 
	\end{pmatrix}
\end{equation}
 where  $\mathbf{s} = \hbar /2 \{\sigma^x,\sigma^y,\sigma^z\}$ and $\sigma^i$ are $2\times2$ Pauli matrices.

We would like to find a mapping between Eqs.~(\ref{eq: 2x2 Zeeman Hamiltonian of 16x16 spin operator}) and (\ref{eq: 2x2 Zeeman Hamiltonian of effective spin}) that preserves the energy splitting between $|v_1\rangle$ and $|v_2\rangle$ in Eq.~(\ref{eq: 2x2 Zeeman Hamiltonian of 16x16 spin operator}). We find, where $\alpha \in \{x,y,z\}$,
\begin{equation}
	\begin{aligned}
		&\langle v_1 | g_L S^\alpha | v_1 \rangle = \pm \langle \uparrow | g_{eff}^\alpha s^\alpha | \uparrow \rangle, \\
		&\langle v_2 | g_L S^\alpha | v_2 \rangle = \pm \langle \downarrow | g_{eff}^\alpha s^\alpha | \downarrow \rangle, \\
		&\langle v_2 | g_L S^\alpha | v_1 \rangle = \pm \langle \downarrow | g_{eff}^\alpha s^\alpha | \uparrow \rangle.
	\end{aligned}
\end{equation}
The effective $g$~factor,
\begin{equation}
	\begin{aligned}
		g_{eff}^\alpha = \pm
		g_L \frac{\langle v_1 | S^\alpha | v_1 	\rangle}{ \langle \uparrow |s^\alpha | \uparrow \rangle},
	\end{aligned}
\end{equation}
where % $p^\alpha$ is defined 
\begin{equation} 
	\begin{aligned}
			p^\alpha = \pm \frac{g_{eff}^\alpha}{g_L} 
			=\frac{\langle v_1 | S^\alpha | v_1 \rangle}{ \langle \uparrow |s^\alpha | \uparrow \rangle}.
	\end{aligned} 
\end{equation}

Within the lowest doublet the  $16\times 16$ spin operator and the 16 states can be replaced by a $2\times 2$ spin operator and the up and down states. The  easy axis and the crystal field are encoded within the $p^\alpha$. The sign of $p$ determines the  direction of the easy axis, as
\begin{equation} \label{eq: transformation of S_16 to s_2}
	\begin{aligned}
		&S^\alpha \rightarrow p^\alpha s^\alpha \\
		&| v_1 \rangle \rightarrow | \uparrow \rangle
	\end{aligned} 
\end{equation}
This transformation reduces the size of the Hilbert space as  we are now dealing with 2 states at each site instead of 16, and those 2 states have an effective $g$ tensor.

\begin{table}
	\caption{Energy splittings and $g$~factors of $C_2$ sites \cite{Dammak_200301822,Ammerlaan2001-kp}.}%
	%%%  unit conversion: 1 cm-1 = 0.124 meV
	\begin{ruledtabular}
		\begin{tabular}{cll}\label{tab: $C_2$ splitting value}
			\textrm{}&
			\textrm{Exp (cm$^{-1}$)}&
			\textrm{This model (cm$^{-1}$)} \\
			\hline\hline
			 &  0 & 0\\
			 &  36& 48\\
			 & 77 & 80\\
			 & 90 & 85\\
			CF splitting & 163 & 149\\
			 & 258 & 272\\
			 & 492 &412\\
			 & 507 & 495\\
			\hline
			$g_x$ & 1.6 & 1.9 \\
			$g_y$ & 4.9 & 5.1 \\
			$g_z$ & 12.3 & 12.6  
		\end{tabular}
	\end{ruledtabular}
\end{table}

\begin{table}
	\caption{Energy splittings and $g$~factors of $C_{3i}$ sites\cite{Dammak_200301822,Ammerlaan2001-kp}.}%
	%%%  unit conversion: 1 cm-1 = 0.124 meV
	\begin{ruledtabular}
		\begin{tabular}{cll}\label{tab: $C_{3i}$ splitting values}
			\textrm{}&
			\textrm{Exp (cm$^{-1}$)}&
			\textrm{This model (cm$^{-1}$)} \\
			\hline\hline
			&  0 & 0\\
			& 41 & 19\\
			& 80 & 67\\
			CF splitting & 328 & 228\\
			& 359 & 260\\
			& 391 & 416\\
			& 416 & 496\\
			& 485 & 542\\
			\hline
			$g_{||}$ & 12.3 & 12.38 \\
			$g_{\perp}$ & 3.3 & 3.38 
		\end{tabular}   
	\end{ruledtabular}
\end{table}

Now that we have constructed a model for the crystal field of erbium at the two inequivalent sites in Er$_2$O$_3$, we proceed to find the symmetry-dependent exchange constants. 

\section{Effective exchange and dipolar constants} \label{sec: effective constants}

Using the above transformation from the $16x16$ Hamiltonian to the $2\times 2$ effective Hamiltonian for the ground-state Kramers doublet and substituting those effective spins into the exchange Hamiltonian yields
\begin{equation}
	\begin{aligned}
		\mathcal{H}_{ex} = &\sum_{(i,j)} J_o  %\\
		%
		%&
		\Big( |v1_i,v1_j \rangle \langle v1_i,v1_j| + |v2_i v2_j\rangle \langle v2_i v2_j| \Big)  \\
		&\qquad \times \Big( S_i^x S_j^x + S_i^y S_j^y + S_i^z S_j^z \Big)\\
		&\times \Big( |v1_i,v1_j \rangle \langle v1_i,v1_j| + |v2_i v2_j\rangle \langle v2_i v2_j| \Big)
	\end{aligned}
\end{equation}
or equivalently
\begin{equation}
	\begin{aligned}
		\mathcal{H}_{ex} = & \sum_{(i,j)} J_o \Big( |\uparrow_i,\uparrow_j \rangle \langle \uparrow_i,\uparrow_j| + |\downarrow_i \downarrow_j\rangle \langle \downarrow_i \downarrow_j| \Big) \\
		&\qquad\times \Big( p^x_i p^x_j s_i^x s_j^x + p^y_i p^y_j  s_i^y s_j^y + p^z_i p^z_j  s_i^z s_j^z \Big)\\
		&\qquad\times \Big( |\uparrow_i,\uparrow_j \rangle \langle \uparrow_i,\uparrow_j| + |\downarrow_i \downarrow_j\rangle \langle \downarrow_i \downarrow_j | \Big)
	\end{aligned}
\end{equation}
For
\begin{equation}
	\begin{aligned}
		&J^\alpha_{(i,j) } = p^\alpha_i p^\alpha_j
	\end{aligned}
\end{equation}
in the $|\uparrow\rangle$ and $|\downarrow\rangle$ basis the exchange Hamiltonian becomes
\begin{equation} 
	\begin{aligned}
		H_{ex} = \sum_{(i,j)} J_o \Big(J^x_{(i,j)} s^x_i s^x_j + J^y_{(i,j)} s^y_i s^y_j+J^z_{(i,j)} s^z_i s^z_j\Big)
	\end{aligned} 
\end{equation}
where $J_o J^z_{C_{2z},C_{2\bar{z}}}$ can be found
within  the molecular field approximation, fixing the value of $J_o$.

For Er$_2$O$_3$ the Ne\'el temperature ($T_N$) is $3.3$~K\cite{NARANG2014353} and the erbium site's six neighbors are oriented anti-parallel. 
If each sublattice is in an Antiferromagnetic-Antiferromagnetic-Antiferromagnetic (AAA) configuration the other terms of the above expression are zero because the easy axes of each neighbor (other than anti-parallel ones located at 5.3 \r{A}) are pointed in opposite directions \cite{MagnetismBook}. This determines the value of $J_o$:
\begin{equation} 
	\begin{aligned}
		J_o J^z_{C_{2z},C_{2\bar{z}}} &= 1.1 (8.62 \times 10^{-2}) {\rm meV} = 0.095 {\rm meV}  \\
		J_o &=  0.095 {\rm meV}  / J^z_{C_{2z},C_{2\bar{z}}} \\
		J_o &=  0.095 {\rm meV}  / (p^z_{C_{2z}})^2\\
		J_o &=  0.095 {\rm meV}  / (10.25)^2\\
		J_o &= 0.9 \mu{\rm eV}
	\end{aligned} 
\end{equation}
%\begin{table}\label{tab: values of p's}
%	\caption{The p's from describing the lowest Kramers doublet as an effective spin-1/2.}%
%	\begin{ruledtabular}
%		\begin{tabular}{|cll|}
%			\textrm{}&
%			\textrm{$C_{2z}$}&
%			\textrm{$C_{3i}$} \\
%			\colrule
%			 $p_x$ & 1.3 &5.89\\
%			 $p_y$ & 4.1&5.89\\
%			 $p_z$ & 10.3&5.89
%		\end{tabular}
%	\end{ruledtabular}
%\end{table}
The coefficient of the dipolar interaction, $D$, is
\begin{equation} 
	\begin{aligned}
		D &= \Big(\frac{-g_L \mu_B}{\hbar}\Big)^2\frac{\mu_o}{4 \pi |\textbf{r}|^3}
	\end{aligned} 
\end{equation}

\noindent where

\begin{equation} 
	\begin{aligned}
		\Big(g_L \mu_B\Big)^2 \frac{\mu_o}{4 \pi} 
		&= 
		0.07727616 \;\; {\rm meV} \; \text{\r{A}}^3
	\end{aligned} 
\end{equation}	

These constants are used  for the calculation of  magnonic dispersion. The next section establishes the formalism for calculating  magnons in this noncollinear system, where the quantization axes vary from site to site, and are not orthogonal.

\section{Noncollinear magnons using Holstein-Primakoff transformation} \label{sec: non collinear magnons}

We use the  coordinate transformation  from Ref.~\onlinecite{sym14081716} to describe the exchange interaction.
The zeroth order Holstein-Primakoff (HP) transformation along with the reduction of the Hilbert space defined by Eq.~(\ref{eq: transformation of S_16 to s_2})  is, with $s_{HP} = (s^1,s^2,s^3)$,
\begin{equation} 
	\begin{aligned}
		&s^1 = p^1 \frac{\sqrt{2s}}{2}(b^\dagger + b)\\
		&s^2 = p^2 \frac{\sqrt{2s}}{2i}(b^\dagger-b)\\
		&s^3 = p^3 (s - b^\dagger b) 
	\end{aligned} 
\end{equation}
Note that $s = 1/2$ because we perform the HP transformation on the effective spin operators.
Let $R$ be a matrix that maps $\hat{z}$ to the easy axis of the site. In the $e$ coordinate system $(\hat{x},\hat{y},\hat{z})$, the components of spin are
\begin{equation} \label{eq: definition of S_sym}
	\begin{aligned}
		\textbf{s}_{sym} = R_{sym}
		\begin{pmatrix}
			S^1\\
			S^2\\
			S^3
		\end{pmatrix} = R_{sym} \textbf{s}_{HP}
	\end{aligned} 
\end{equation} 
This $\textbf{s}_{sym} $ is in the $e$ coordinate system and $R_{sym}$ is a rotation matrix that depends on the local site symmetry. These rotation matrices are 
\begin{equation} 
	\begin{aligned}
		R_{-\hat{z}} =&\ diag(1,-1,-1),\\
		R_{\hat{x}} =& 
		\begin{pmatrix}
			0&0&1\\
			1&0&0\\
			0&1&0
		\end{pmatrix},\\
		 R_{-\hat{x}} =& 
		\begin{pmatrix}
			0&0&-1\\
			1&0&0\\
			0&-1&0
		\end{pmatrix},\\ 
		R_{\hat{y}} = &
		\begin{pmatrix}
			0&1&0\\
			0&0&1\\
			1&0&0		 	
		\end{pmatrix},\\
		R_{-\hat{y}} =& 
		\begin{pmatrix}
			0&-1&0\\
			0&0&-1\\
			1&0&0
		\end{pmatrix}. 
	\end{aligned} 
\end{equation}
and the rotation matrix for the $C_{3i}$ site is, with $c = \cos (\theta)$ and $s = \sin(\theta)$,
\begin{equation} \label{eq: rotation matrix of $C_{3i}$}
	\begin{aligned}
		R(\theta) = \begin{pmatrix}
			\frac{1}{2} (1+ c)&\epsilon\frac{1}{2}  (-1+ c) &\frac{s}{\sqrt{2}}\\
			\epsilon\frac{1}{2} (-1+ c)&\frac{1}{2} (1+ c)&\epsilon\frac{s}{\sqrt{2}}\\
			\frac{-s}{\sqrt{2}}&\epsilon\frac{- s}{\sqrt{2}}&c
		\end{pmatrix}.
	\end{aligned} 
\end{equation}
The parameters of this matrix ($\epsilon$ and $\theta$) are in Table~\ref{tab: matrix details}.  
%One can simply get these matrices using wolfram alpha RotationMatrix[$\theta$,$v_\epsilon$]), where $\theta$ is given in Tab. \ref{tab: matrix details}. $v_{+1}$ and $v_{-1}$ are $(-1,1,0)$ and $(1,1,0)$, respectively.
\begin{table}
	\caption{Values of $\epsilon$ and $\theta$  in the definition of the rotation matrix of $C_{3i}$ sites, Eq. (\ref{eq: rotation matrix of $C_{3i}$}).}%
	\begin{ruledtabular}\label{tab: matrix details}
		\begin{tabular}{|cll|}
			\textrm{symmetry axis}&
			\textrm{$\epsilon$}&
			\textrm{$\theta$}  \\
			\colrule
			$[111]$ & +1  & $\omega$\footnote{$\omega = \cos^{-1}(1/\sqrt{3})$}\\
			$[\bar{1}\bar{1}1]$ & $+1$ &$-\omega$\\
			$[1\bar{1}1]$ & $-1$ & $\omega$ \\
			$[\bar{1}11]$ & $-1$ & $-\omega$ \\
			$[11\bar{1}]$ & $+1$ & $-\omega - \pi$ \\
			$[\bar{1}\bar{1}\bar{1}]$ & $+1$ & $\omega - \pi$ \\
			$[\bar{1}1\bar{1}]$ & $-1$ & $\omega - \pi$ \\
			$[1\bar{1}\bar{1}]$ & $-1$ & $-\omega - \pi$ \\
		\end{tabular}
	\end{ruledtabular}
\end{table}
For example, the interaction between $C_{2z}$ and $C_{2x}$,
\begin{equation} 
	\begin{aligned}
		\textbf{s}_{i, C_{2z}} \cdot \textbf{s}_{j,C_{2x}} &= \textbf{s}_{i,HP} R^T_{C_{2z}} R_{C_{2x}} \textbf{s}_{j,HP}.
	\end{aligned} 
\end{equation}
Here
$s_{sym}$ defined in Eq.~(\ref{eq: definition of S_sym}) can also be used in the dipolar Hamiltonian because both $\hat{r}$ (the unit vector connecting the two sites) and $s_{sym}$ are in the $e$ coordinate system.

The Hamiltonian which governs the dynamics of both exchange and dipolar magnons is
\begin{equation} \label{eq: Ham = H_dipolar + H_ex + H_ext }
	\begin{aligned}
		\mathcal{H} &= \mathcal{H}_d + \mathcal{H}_e + \mathcal{H}_{ext}\\
		&=
		 \sum_{<i,j>} \Biggl\{D \Big(3 (\textbf{S}_i\cdot \hat{r})(\textbf{S}_j\cdot \hat{r})\Big) - D \textbf{S}_i\cdot \textbf{S}_j - J_o \textbf{S}_i\cdot \textbf{S}_j
		\Biggr\} \\
		&\qquad\qquad- 
		\sum_i \frac{g_L \mu_B}{\hbar} \textbf{S}_{i} \cdot \textbf{B}_{ext} .
	\end{aligned} 
\end{equation}
After performing the HP and Hilbert space reduction we obtain a bilinear Hamiltonian in terms of magnon operators which has the form
\begin{equation} 
	\begin{aligned}
		\mathcal{H} &= 
		\sum_{<i,j>} \Big(\beta_1 b_i b_j + \beta_1^\ast b_i^\dagger b_j^\dagger + \beta_2 b_i^\dagger b_j + \beta_2^\ast b_i b_j^\dagger +  \\
		&\qquad\qquad+ 
		\beta_3 b_i^\dagger b_i + \beta_4 b_j^\dagger b_j
		\Big) 	
	\end{aligned} 
\end{equation}
where $\beta$'s are functions of the distance between the sites, the symmetry of each site, the external $B$ field, and the exchange and dipolar constants. 
In order to diagonalize this Hamiltonian, we first write it in a more convenient form,
\begin{equation}
	2\frac{\mathcal{H}\,}{s}=\,\left(b_{1}^{\dagger}\,...b_{32}^{\dagger}\,b_{1}...b_{32}\right)\begin{pmatrix}
		{H_{11}}&{H_{12}}\\
		{H_{21}}&{H_{22}}\\
	\end{pmatrix}\begin{pmatrix}
		{b_{1}}\\
		{...}\\
		{b_{32}}\\
		{b_{1}^{\dagger}}\\
		{...}\\
		{b_{32}^{\dagger}}\\
	\end{pmatrix}=\psi^{\dagger}M\psi.
\end{equation}

Diagonalization of $M$ is not sufficient because it will not conserve the bosonic commutation relationship of the creation and annihilation operators. This point can be clarified if we look at the equations of motion of  $b$ and $b^\dagger$,
\begin{eqnarray}
	i \frac{d}{dt} b &&= H_{11} b + H_{12} b^\dagger \\
	i \frac{d}{dt} b^\dagger &&= -H_{21} b - H_{22} b^\dagger,
\end{eqnarray}
The negative sign on the right hand side of this equation of motion for $b^\dagger$ forces us to diagonalize it paraunitarily using \cite{Xiao2009TheoryOT} 
\begin{equation}\label{eq: para definition}
	|H-\omega \mathcal{I}_{-}  |=0,
\end{equation}
where % $\mathcal{I}_{-}$ 
\begin{equation}
	\mathcal{I}_{-} = \begin{pmatrix}
		\mathcal{I} & 0 \\ 0 & -\mathcal{I} \\ 
	\end{pmatrix}.
\end{equation}

We multiply the eigenvalue equation by $\mathcal{I}_{-}$  to obtain
\begin{equation}\label{eq: para definition with D}
	|\mathcal{D}-\omega \mathcal{I}  |=0
\end{equation}
where $\mathcal{D} = H \mathcal{I}_{-}$ is non-Hermitian. This form of diagonalization guarantees particle-antiparticle symmetry, but it does not always yield real eigenvalues. However, if our ground state is chosen correctly and the oscillations around the ground state are small the eigenvalues are real. For a small external magnetic field the eigenvalues shift slightly but if the applied external field is too large a  spin flop can occur within the system and the magnon dispersion will yield complex eigenvalues. This would indicate that the correct equilibrium configuration for the sites is not the initially assumed  configuration.

\section{Magnon dispersion}\label{sec: Magnon dispersion}
	
The exchange interaction drops exponentially with distance whereas the dipolar interaction drops as a cube of the distance between the two sites. We only consider the exchange interaction between  sites  closer than $5.3$ \r{A},  corresponding to the first three nearest neighbors. Up to the third nearest neighbor are included as the strongest exchange interaction will occur between  anti-parallel sites, and the first anti-parallel neighbor (that is, the first neighbor with the same quantization axis as at the origin) is the third nearest neighbor located at $5.3$ \r{A}. For the dipolar interaction, study the change in dispersion curves as the number of included neighbors is increased, as the decay of the interaction is far slower than exchange. An analysis of the convergence of these dispersion curves as the number of the dipolar neighbors increases suggests that many neighbors should be included, however the eventually the results are robust to the addition of additional neighbors. 

\begin{figure} [h!]
	% \begin{figure*} makes it one column 
		\centering
		\includegraphics[width = 0.4\textwidth]{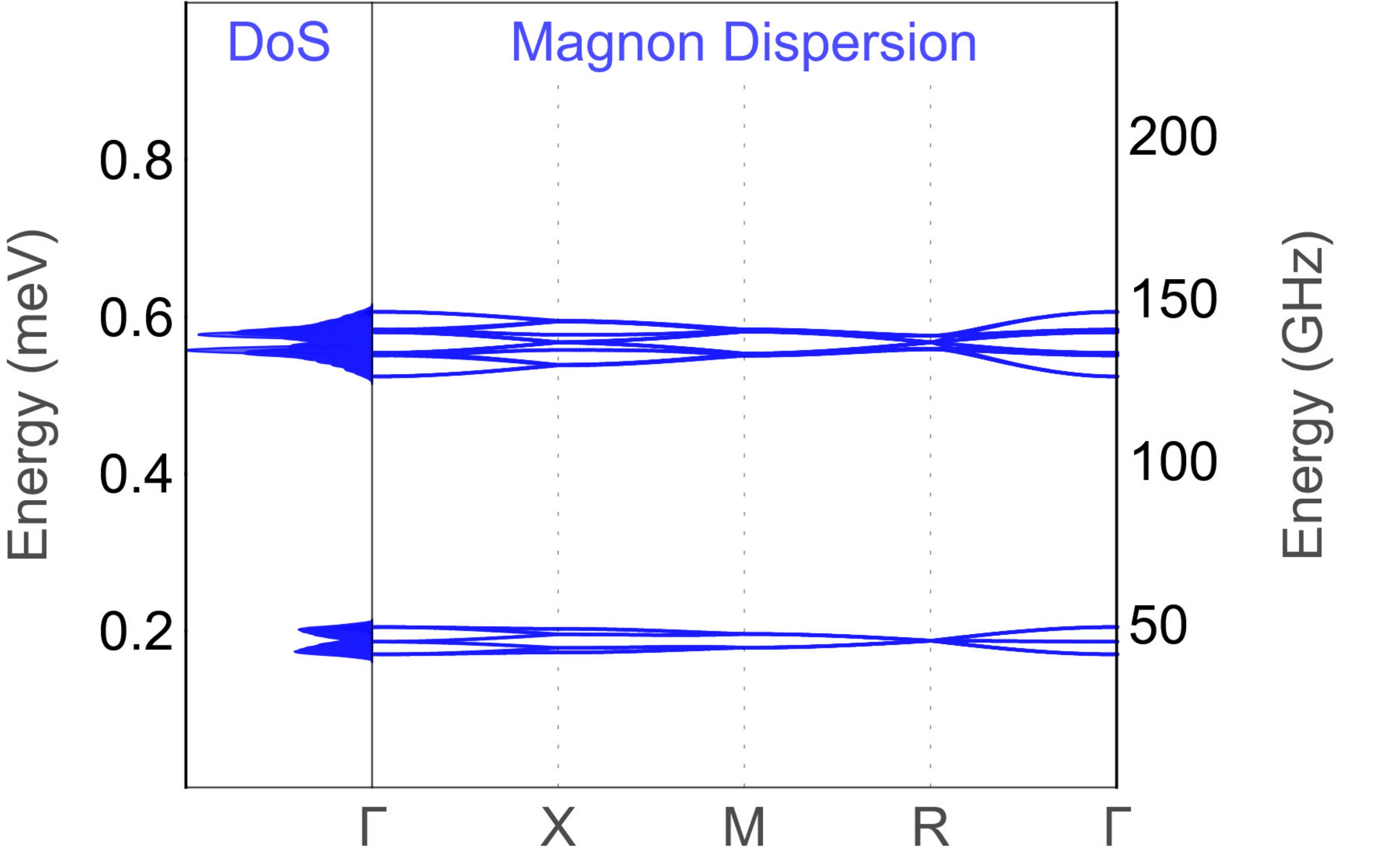}
		\caption{Magnon dispersion at zero applied magnetic field due to the first three nearest neighbor exchange interactions (no dipolar interactions). The upper bands originate from interactions among the antiparallel $C_2$ sites, and the lower bands originate from the other interactions.}
		\label{fig: dispersion exhcnage only}
\end{figure}

\begin{figure} 
	% \begin{figure*} makes it one column 
    \centering
  \begin{annotate}{\includegraphics[width=0.39\textwidth]{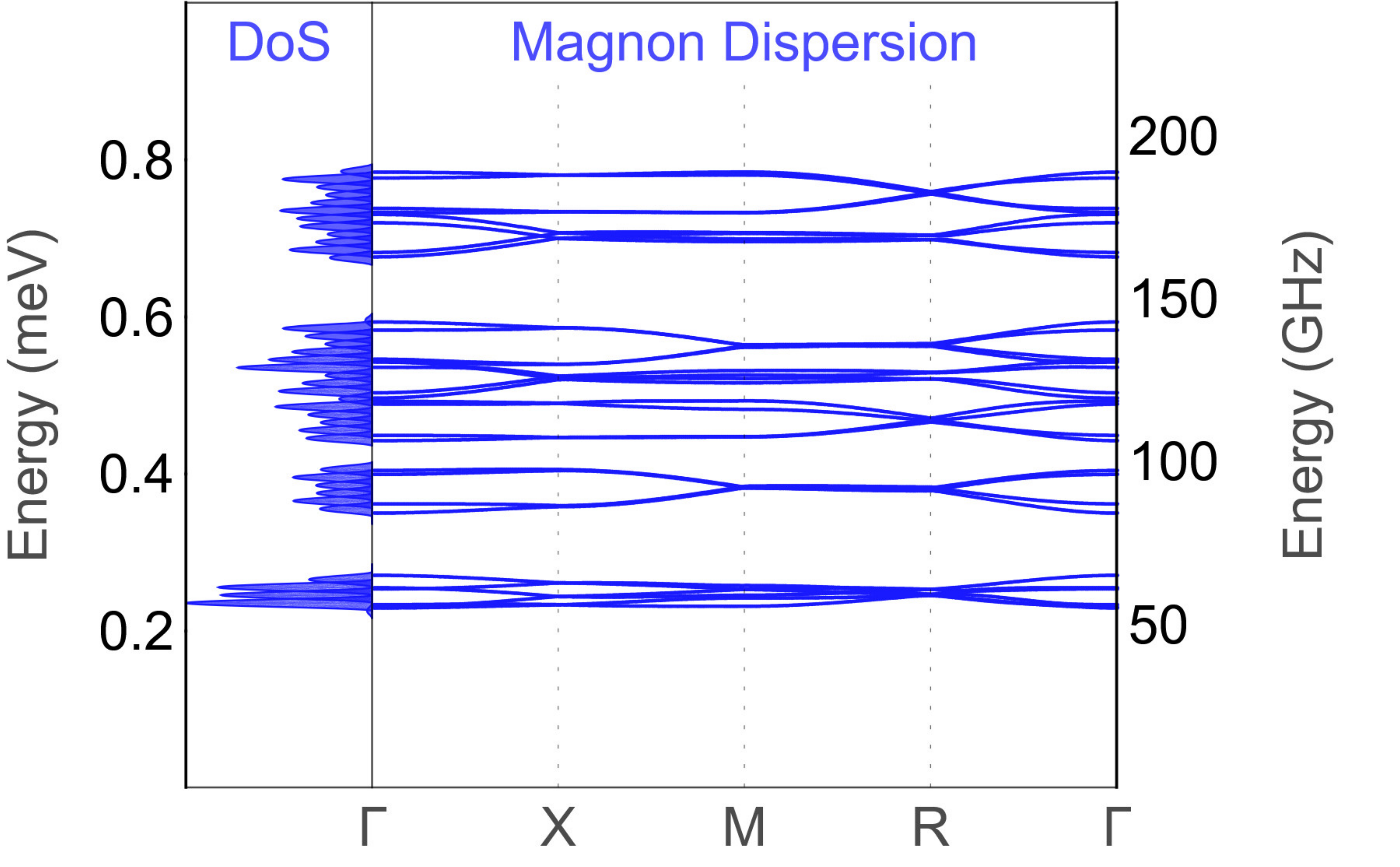}}{1}
    \note{-3.5,2}{a)}
   \end{annotate}
  \begin{annotate}{\includegraphics[width=0.39\textwidth]{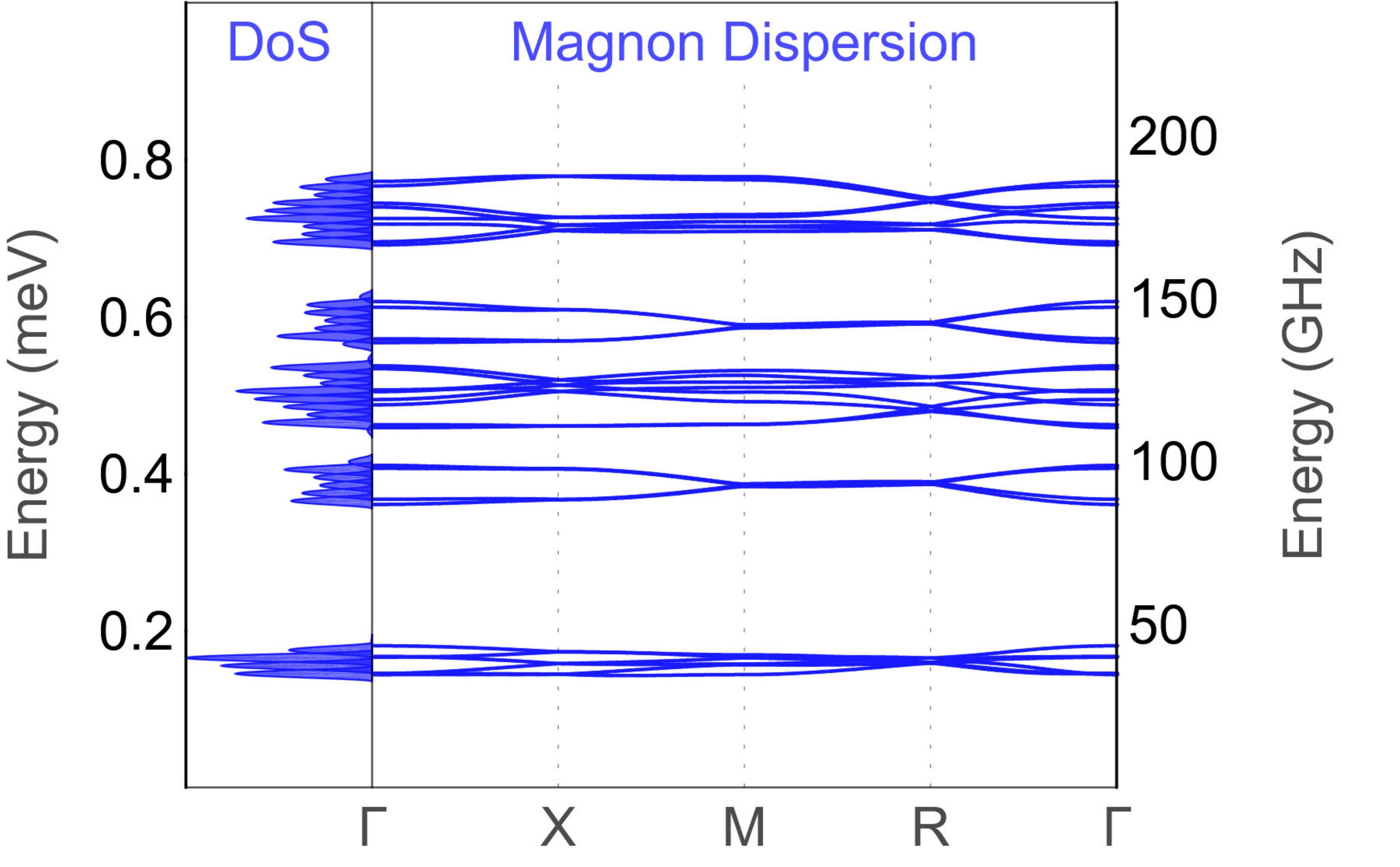}}{1}
    \note{-3.5,2}{b)}
   \end{annotate}
  \begin{annotate}{\includegraphics[width=0.39\textwidth]{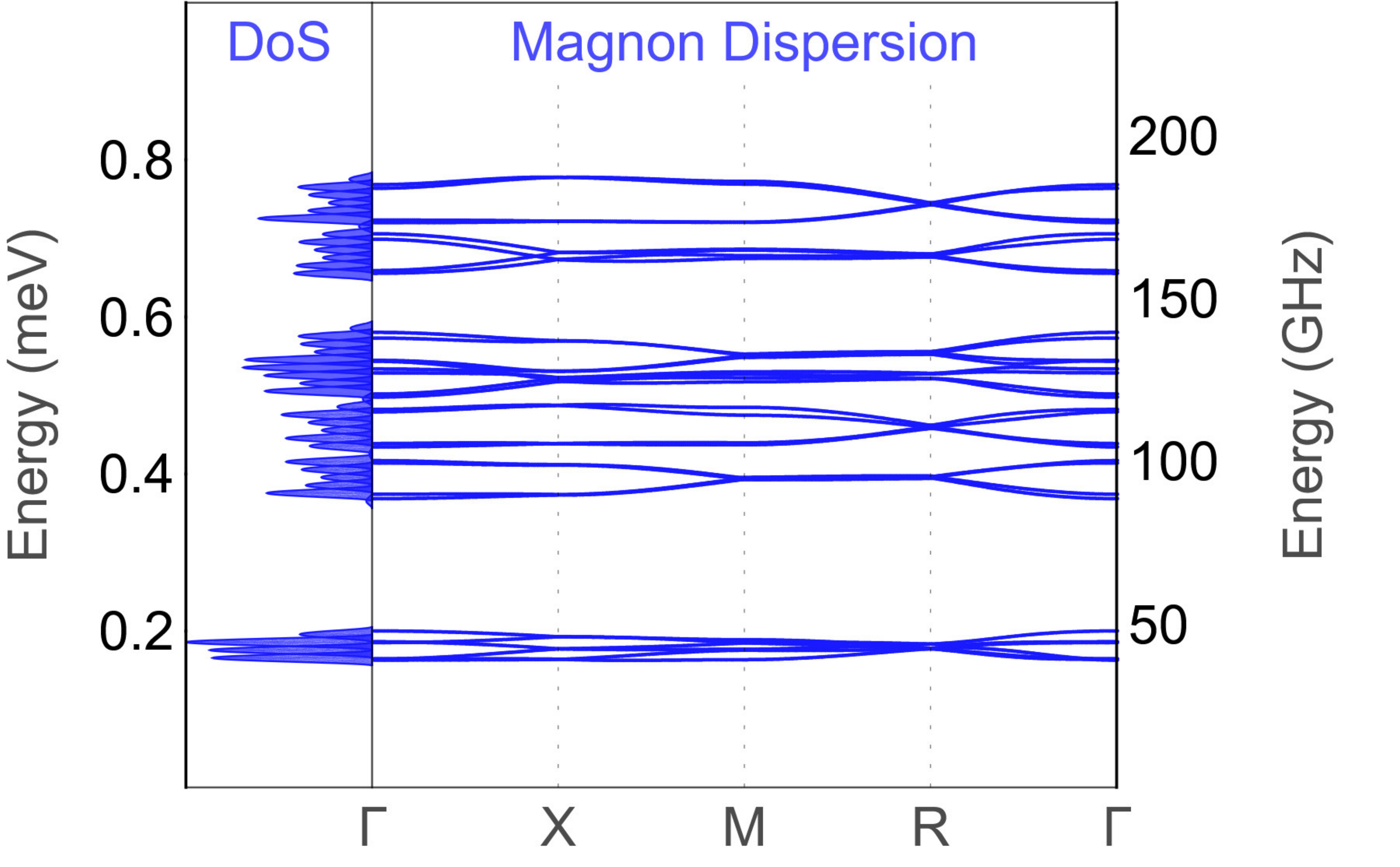}}{1}
    \note{-3.5,2}{c)}
   \end{annotate}
     \begin{annotate}{\includegraphics[width=0.39\textwidth]{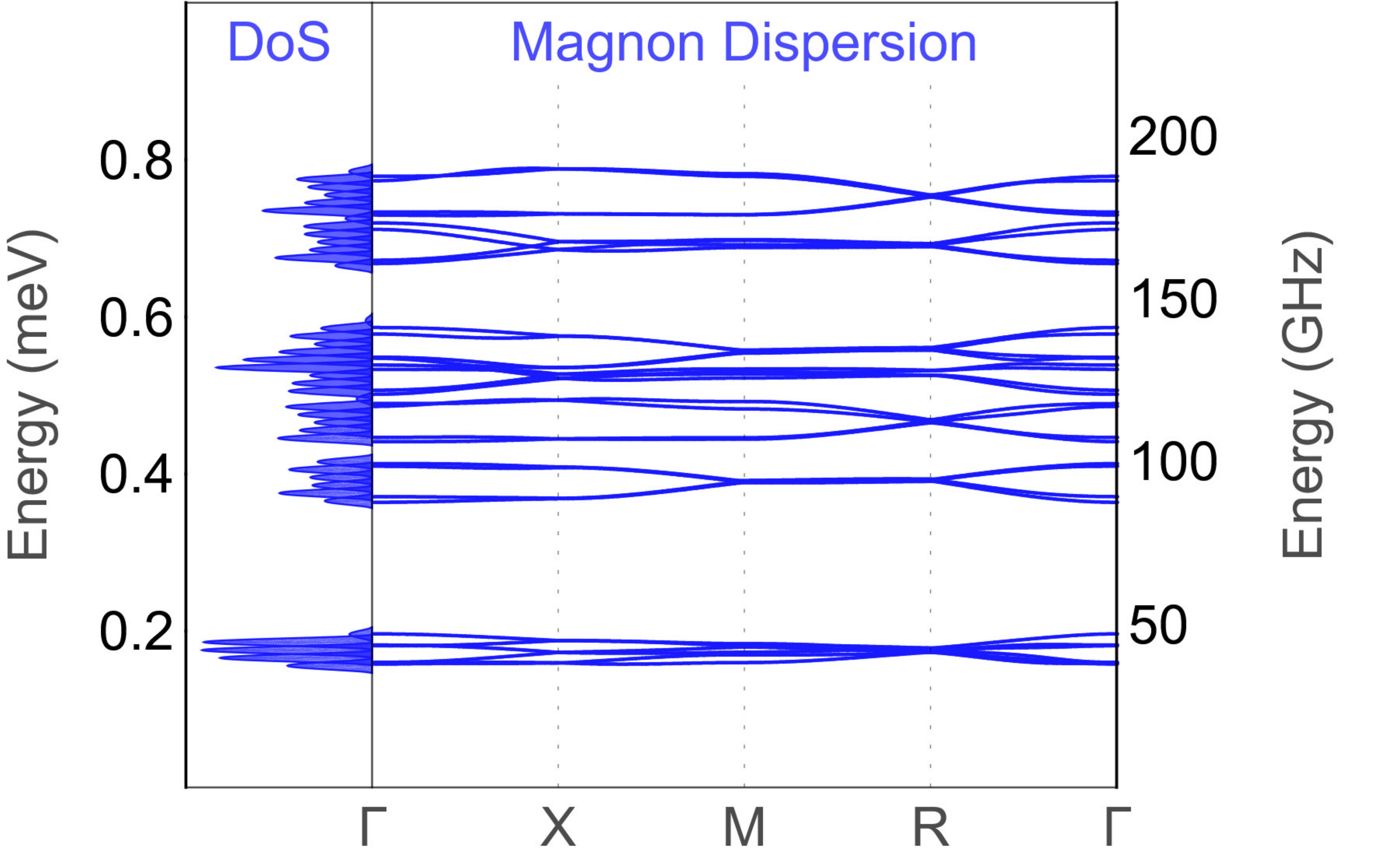}}{1}
    \note{-3.5,2}{d)}
   \end{annotate}   
		\caption{Magnon dispersion in the presence of exchange and dipolar interactions for twenty, forty, sixty and eighty dipolar nearest-neighbor interactions,  in panels (a), (b), (c), and (d).  }
		\label{fig: dispersion dipolar ALL 4 of them}	
\end{figure}

\begin{figure}
	% \begin{figure*} makes it one column 
		\centering
		\includegraphics[width = 0.45\textwidth]{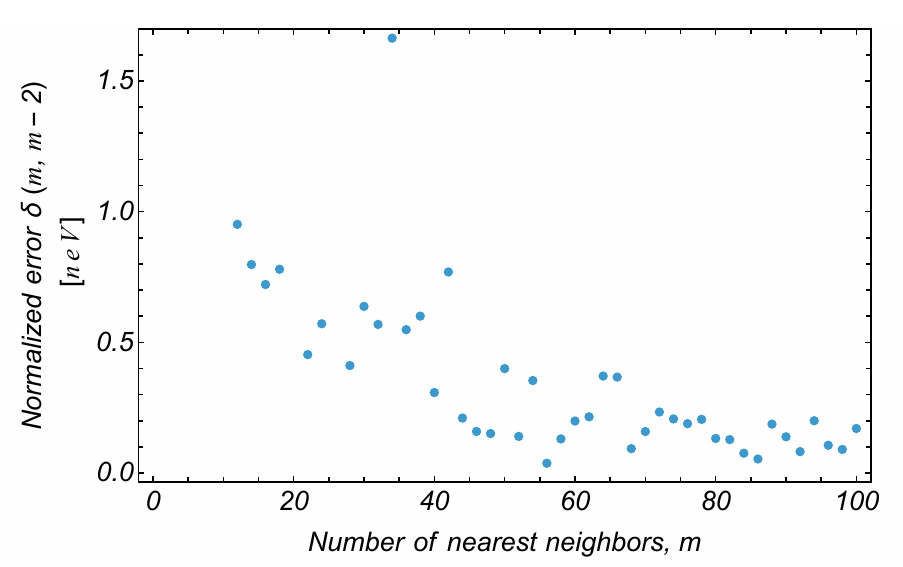}
		\caption{The  error $\delta(m,m-2)$ from  Eq.~(\ref{eq: relative error}) is plotted here  vs $m$.  }
		\label{fig: convergence}
\end{figure}

The magnon dispersion  with only exchange interactions included is shown in Fig.~\ref{fig: dispersion exhcnage only}. The upper bands originate from the interaction among  $C_2$ sites which are anti-parallel, and the lower bands originate from the rest of the exchange interactions.
If the dipolar interaction is included, as shown in 
Fig.~\ref{fig: dispersion dipolar ALL 4 of them}, the bands straighten out somewhat but are not dramatically changed. Fig.~\ref{fig: dispersion dipolar ALL 4 of them} shows the magnon dispersion curve when the first twenty, forty, sixty, and eighty nearest neighbors, by distance, are included in the dipolar interaction. As the dipolar interaction is long-range the magnon dispersion is much less sensitive to the detailed structure of the unit cell.
Not much difference is directly evident between the dispersion with forty and eighty nearest neighbors interaction via dipolar interactions. In order to quantify the trends with additional neighbors we consider a normalized root mean square difference between the dispersion curves, 
\begin{equation}\label{eq: relative error}
	\delta(m,m-2) = \frac{1}{N k} \sqrt{\sum^{N}_{n=1} \Big(\omega^m_n-\omega^{m-2}_n\Big)^2},
\end{equation}
where $N$ is the total number of points on the horizontal axis of each dispersion plot. For our calculations here there $N=19200$, because there are 600 values on the horizontal axis and at each of these values there are 32 magnon frequencies. 
%$n$ runs over all of the data points on x the axis. 
$m$ is the number of nearest neighbors that are used in the calculation of the dipolar dispersion.
$k$ is the total number of pairwise Er-Er interactions that are present in the calculation of the dispersion for $m$ neighbors but not for $m-2$ neighbors. For example, for $m=100$ the value of $k$ is 480.
%$\delta (m,m-2)$ is what we call the normalized and average difference between dispersion with $m$ and $m-2$ dipolar nearest neighbors.
Fig. \ref{fig: convergence} shows $\delta(m,m-2) $ as a function of $m$, indicating convergence. For example, $\delta(m,m-1)<0.5$~neV for $m\ge 40$. As the approximate average value of the magnon frequency in these dispersions is about 0.5~meV, this places the relative error at about $10^{-6}$.

\begin{figure} 
	% \begin{figure*} makes it one column 
		\centering
     \begin{annotate}{\includegraphics[width=0.39\textwidth]{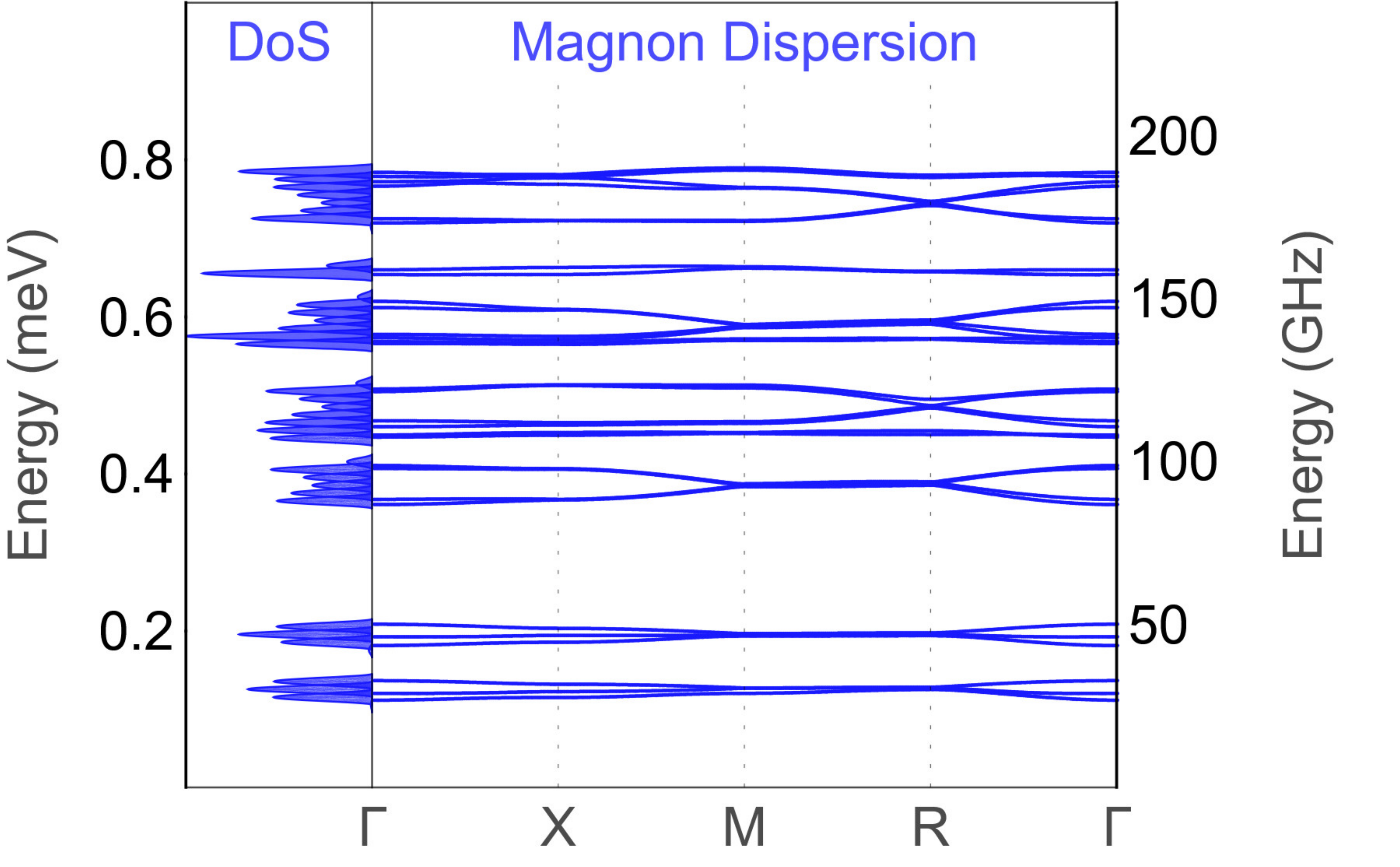}}{1}
    \note{-3.5,2}{a)}
   \end{annotate} 
     \begin{annotate}{\includegraphics[width=0.39\textwidth]{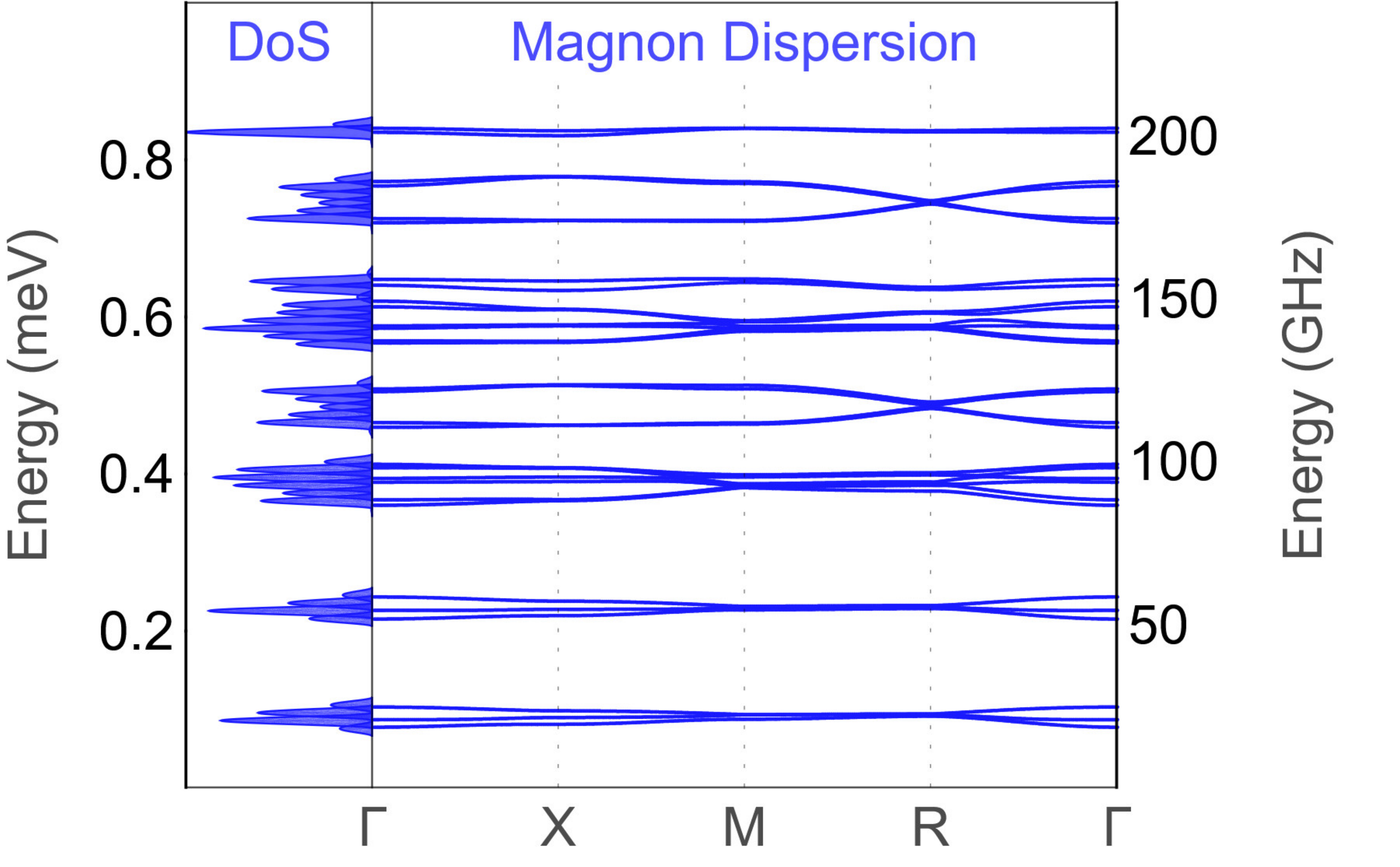}}{1}
    \note{-3.5,2}{b)}
   \end{annotate} 
     \begin{annotate}{\includegraphics[width=0.39\textwidth]{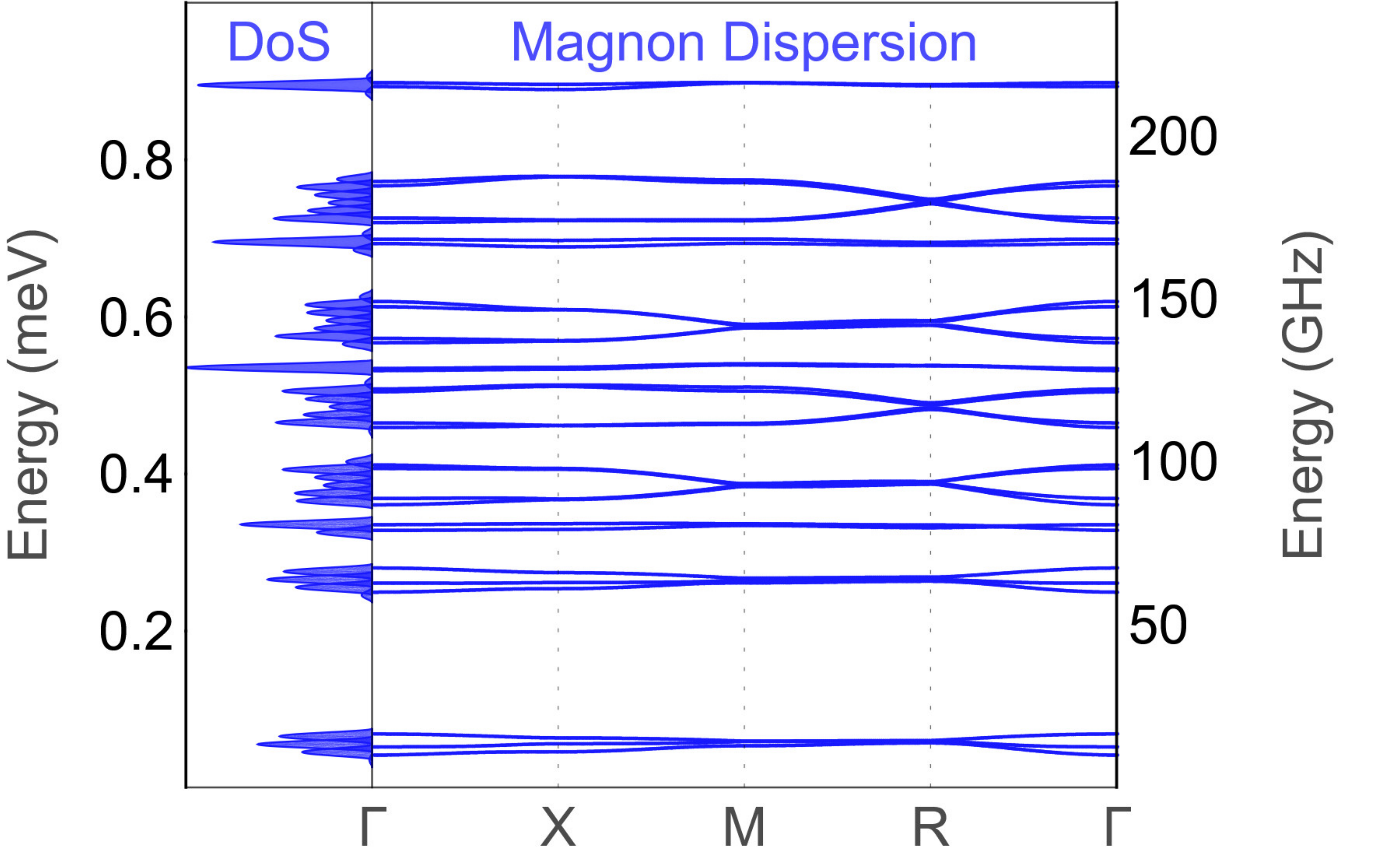}}{1}
    \note{-3.5,2}{c)}
   \end{annotate} 
		\caption{Same as above with forty dipolar nearest neighbors, but with an external magnetic field.  (a) 0.1 T, (b) 0.2 T, and (c) 0.3 T. }
		\label{fig: dispersion external}	
\end{figure}

To study the properties in a finite field the number of dipolar nearest neighbors is fixed at forty, and an external magnetic field is introduced along an axis of the cube (parallel to a $C_2$ site symmetry axis). These results are shown in  Fig. \ref{fig: dispersion external}. With an external magnetic field stronger than 0.33~T the ground state undergoes a spin flop transition, invalidating our approach and producing complex magnon frequencies. When spin flop happens the spin of the sites align along the external magnetic field. % which may be along their hard axis. 
 
\section{Conclusion}
We have used Steven's operators to parametrize the crystal field in Er$_2$O$_3$. The free parameters of this model were fit to the crustal field and $g$~factors. We then used the effective $g$~factors to calculate the symmetry dependent exchange interaction. In the magnon calculations both exchange and  dipolar interactions were included. The long range nature of the dipolar interaction poses a challenge to keep sufficient relevant neighbors in the magnon dispersion. Forty nearest neighbors appear sufficient for the calculations presented here. If the external magnetic field is larger than 0.33~T evidence of a spin flop is seen in the calculations. 

\begin{acknowledgments}
This work was supported by the U.S. Department of
Energy, Office of Science, Office of Basic Energy Sciences, Division of Materials Sciences and Engineering 
under Award No. DE-SC0023393.  We acknowledge useful conversations with D. D. Awschalom, L. Bassett, A. Faraon, D. A. Fehr, J. Lizarazo, T. O. Puel, J. Thompson, and  T. Zhong.
\end{acknowledgments}

\end{document}